\documentclass[11pt]{article}
\usepackage{amsfonts}
\usepackage{amsmath}
\usepackage{natbib}
\usepackage{hyperref}
\usepackage{graphicx}
\usepackage[margin=4.0cm]{geometry}
\usepackage{multirow}
\DeclareMathOperator*{\argmin}{arg\,min}

\DeclareMathOperator*{\var}{\mathrm{var}}

\title{A general decision framework for structuring computation using Data Directional Scaling to process massive similarity matrices} 

\author{Daniel John Lawson\thanks{Heilbronn Institute, School of Mathematics, University of Bristol}\hspace{0.2cm} and Niall Adams\footnotemark[1] \thanks{Department of Mathematics, Imperial College London}}
\date{\today}
\begin{document}
\maketitle
\begin{abstract}
As datasets grow it becomes infeasible to process them completely with a desired model. For giant datasets, we frame the order in which computation is performed as a decision problem. The order is designed so that partial computations are of value and early stopping yields useful results. Our approach comprises two related tools: a decision framework to choose the order to perform computations, and an emulation framework to enable estimation of the unevaluated computations. The approach is applied to the problem of computing similarity matrices, for which the cost of computation grows quadratically with the number of objects. Reasoning about similarities before they are observed introduces difficulties as there is no natural space and hence comparisons are difficult. We solve this by introducing a computationally convenient form of multidimensional scaling we call `data directional scaling'. High quality estimation is possible with massively reduced computation from the naive approach, and can be scaled to very large matrices. The approach is applied to the practical problem of assessing genetic similarity in population genetics. The use of statistical reasoning in decision making for large scale problems promises to be an important tool in applying statistical methodology to Big Data.
\end{abstract}

\section{Introduction}

If a dataset is too large to naively process with a desired model, we can either a) change the model, b) discard some data, or c) selectively order the calculation to yield useful partial results.  This paper describes a framework for decision making to approximately evaluate complex calculations, with application to the specific problem of computing distance or more general similarity matrices.  Statistical reasoning is used to navigate the computational challenges involved. 

The framework has the character of a sequential decision
problem. Specifically, at any sequential step a set of objects are
selected for evaluation. The determination of which objects to
evaluate is guided by a loss function. This loss function cannot be evaluated without the objects, and so we use an \emph{emulator} to enable estimation of the loss. This emulator attempts to provide a computationally efficient prediction of unevaluated objects.  To be of value, this computation must be less demanding than direct evaluation. 

The general framework is applied to the problem of computing similarity matrices.
There are numerous uses of similarity matrices in data
analysis, including clustering, near neighbour search, and anomaly
detection. In principle, evaluating the entire similarity matrix
provides a solution for these uses.  The current focus on ``big
data'', fuelled by technology capable of automatically collecting and
storing data at huge scale, provides new challenges for data
analysis. Since the number of elements in a similarity matrix is
proportional to the square of the number of objects, evaluating the
entire matrix soon becomes computationally intractable. Additionally, we may not always require the complete evaluation of the similarity matrix. This paper develops
a framework for a variety of data analysis activities based on
similarity matrices that is intended to partially address the
computational challenges of big data. 

Data analytic uses of similarity matrices are used in numerous
areas. Bioinformatics is one, with applications including
sequence alignment \cite{thompson1994clustal}, and population genetics \cite{lawson2012population}.  A particularly topical driver is recommender systems
(eg. \cite{rec-sys-book}), which has motivated much work in the
relatively new field of matrix completion
(e.g. \cite{candes2010matrix}). This field  provides tools to impute the missing entries in a matrix. Whilst this is a very useful tool for computation, it is possible to make progress with less computationally intensive approaches.
%This is discussed in more detail below, but the computational effort would not be justified if only part of the matrix was required. 

%It is frequently unnecessary to complete a computation on the large dataset, if the result is going to be used in such a way that full precision is unnecessary. Consider the problem of being presented with a very large number $N$ of independent, identically distributed measurements, with the requirement to calculate the mean. If a desired variance $\sigma^2$ of the estimator can also be specified, then it is possible to perform the computation in $O(1/\sigma^2)$ computations, which is $O(1)$.  There are usually many sources of error in the use of a given dataset, of which estimation is only one. If the estimation error does not dominate, as is typically the case for big datasets, then an approximate but rapid answer is often to be preferred. (cite cite).

%Independent data are in some sense difficult because we can never know the value of a variable without looking at it, fundamentally limiting the computational tricks possible. When data are correlated, it is possible to exploit the correlation to make accurate predictions about unobserved data.  We focus here on a particular sort of correlation found in data measured in terms of similarities.

The methodology we propose directly addresses scenarios involving data on $N$
objects, where $N$ is potentially very large. Objects need not be of the same dimension, but a similarity matrix $S_{ij}$ can be constructed via a similarity function $S(i,j)$. A critical feature of these scenarios is that the similarities are available, but
their number and computational cost
make brute force evaluation intractable. This feature is an important
contrast from the standard matrix completion scenario. Further, there is no feature space in which to place objects. Hence we use the evaluated objects to create a dynamic space via a process we call `data directional scaling'.

Our framework uses statistical reasoning to carefully order a computation to obtain the maximum information benefit for the lowest possible computational cost. To do this, we draw on a number of statistical and machine-learning topics, which are reviewed in Section \ref{sec:lit}.  Our methodology is introduced in Section \ref{sec:method}, including a simple (non-similarity based) example.  We then define the similarity framework itself and introduce our computational framework, along with a simulation study to examine performance as we define the details. Section \ref{sec:scaling} examines how the approach scales with the size of the problem, with Section \ref{sec:genetics} providing a real-world application to a large genetics problem. We conclude in Section \ref{sec:discussion} with a discussion.

\section{Literature review}\label{sec:lit}
This section reviews statistical, machine learning and computer science literature that is helpful or strongly related to our methodology.
Section \ref{sec:emulation} discusses the use of emulators, Section \ref{sec:completion} describes matrix completion, Section \ref{sec:index} describes efficient indexing strategies, Section \ref{sec:active} describes active learning, and Section \ref{sec:sequential} describes sequential decision making.

\subsection{Emulation}\label{sec:emulation}

Gaussian Process emulation \cite{santner2003design} is a major tool in the design and analysis of computer experiments, with applications ranging from climate models \cite{rougier2009analyzing}, and carbon budgets \cite{kennedy2006case} to demography \cite{bijak2013reforging} and testing cars in crashes \cite{bayarri2009predicting}. In typical use, a limited number of computer experiments can be run and the goal is to choose the parameter values to run the expensive simulation model at, in order to fit a costly simulator to data. If the prediction space can be treated as continuous, then treating the observations as varying smoothly according to a multivariate Normal distribution proves surprisingly helpful in predicting good choices for where to evaluate the full simulator.  There are naturally difficulties, one of which being that the cost of estimating a Gaussian Process grows rapidly with the size of the dataset. Decision making to exclude some data for future predictions \cite{seo2000gaussian} can limit this when the data are added sequentially (e.g. \cite{busby2008adaptive}). In this case diagnostics are of extra importance \cite{bastos2009diagnostics}, to prevent the model from exploring the wrong area of the space.

\cite{gramacy2013local} provide a computation-aware framework for Gaussian Processes. Whilst Gaussian Process emulators are by far the most common, other emulators are possible; for example, \cite{vsochman2009learning} use an emulator over a binary space to speed computer vision. Bespoke emulators are also commonplace in the analysis of computer networks (e.g. \cite{simmonds2000applying}). 

Our work differs from the above in a) providing a general class of non-Gaussian Process emulator for similarity matrices; and b) extending the use of emulators to the sort of tradeoff found in everyday computation in `big data'. This is in contrast to the standard emulator framework in which the emulator has negligible computational cost.

\subsection{Matrix Completion} \label{sec:completion}

The very active area of matrix completion is broadly concerned with
imputing missing entries in a matrix from a relatively small
proportion of observed entries.  \cite{candes2010matrix}
provides a summary of recent work on matrix completion. The basic
framework has two key features. First, the observed entries are
assumed to be a random sample and are such that the observed matrix is
full rank. Second, it is assumed that the full matrix has a low rank
representation. Under these assumptions, algorithms implementing nuclear
norm minimisation provide the computational means to attack the matrix
completion problem. Notably proofs are available that a low rank
solution can be recovered in few evaluations within a theoretical
bound, in terms of the root-mean-square reconstruction error. A more
recent review and a novel algorithm working under relaxed
assumptions is provided
by~\cite{CompressedSensingChapter2014}. Indeed, many variations and
refinements have been proposed, including \cite{IEEEtit2010}
which provides an algorithm that claims to open the door for matrix
completion with big data. Scaling to big data, the target of our
proposed framework, is a key challenge in matrix completion. To quote
\cite{candes2010matrix}: ``\ldots but the computational challenges of
solving problems with millions if not billions of unknowns obviously
still require much research''.  The methodology and SOFT IMPUTE algorithm
developed in \cite{mazumder2010spectral} provides capability to handle much
bigger problems than many proposed matrix completion
algorithms. Motivated by the SOFT IMPUTE algorithm and the fact the
big data sometimes features sequential arrival,
\cite{Bhojanapalli2014} develop an online variant of matrix
completion. They key aspect in this approach is a randomised
technique for computing the singular value decomposition - a method
that is at the core of all matrix completion algorithms. 

In addition to the application of matrix completion to recommender
systems discussed in the introduction, other interesting applications
include genotype imputation \cite{Chi01032013}, functional genomics
\cite{gerlee13:_searc_syner}, computer network performance prediction
\cite{liao:_2013} and global positioning in sensor networks
(eg. \cite{biswas2006}). This is an interesting example because
there are often power constraints on sensors that means some distances
are not observed, leading to a bias in the sampling mechanism. This is
explored in \cite{Taghizadeh}.

%% this paper \cite{NIPS2013_4999} tries to use side information to reduce the number
%% of obtained entries. 

%% The issue of some assumptions, notably random sampling, is addressed
%% in \cite{2014arXiv1402.2324B}. Similar guarantees are provided other
%% diffrent sampling schemes.  

%% Similarly, \cite{bernoulli2014} shows general results for sampling,
%% that also do not require the variance of the noise (which is core to
%% the first paper, and ALSO SOMETHING WE ARE HAVING TO DO)

%% The paper ``Coherent Matrix Completion'' has some more specific
%% results.
Matrix completion would be a valid approach to some of the problems we face in this work. However, it is not the approach we take for several reasons. First, we want to make informed decisions with little information, when a matrix completion would be of insufficient rank. Second, the requirement to obtain a full rank solution also provides a restriction on which elements can be evaluated, as all rows and all columns must be visited. Third, little attention has been paid to how to update a matrix completion as data is observed sequentially. Fourth, we can make use of limited information without performing a full matrix completion. Finally, the literature does not provide an `off the shelf' solution tailored to the particular type of similarity matrices we face; as we shall see below, the current methods perform badly because they make the wrong assumptions about the form of the matrix. 

%% This is related to the `matrix completion' problem \cite{ma2011fixed}, in which we are given an incomplete matrix and the goal is to fill in the missing entries. This can be solved efficiently for large matrices \cite{mazumder2010spectral} and it can be shown that this problem is has a good solution \cite{candes2010matrix} when the underlying data have low rank compared to the number of observed entries, the data are sufficiently dense, and the similarities are observed at random.  In our problem, however, we have complete freedom to choose the similarity entries to evaluate, and generally do not believe that the underlying matrix is appropriately dense. We explore some methods of iteratively choosing elements to evaluate, which account for the increasing knowledge of the true underlying structure as the matrix becomes more complete. Our method is capable of fully inferring the whole matrix, and we compare it with competing approaches for matrix completion.

\subsection{Indexing methods} \label{sec:index}

There are numerous data structures useful for indexing multivariate
spaces, including KD-trees \cite{bentley1975multidimensional}, Quadtrees \cite{finkel1974quad} R
trees \cite{Guttman1984Rtrees}, and X-trees \cite{berchtold2001x}. Our problem does not observe a multivariate space, but instead we view similarities as lying on some implicit manifold in an \emph{unknown} space. These approaches therefore cannot be directly applied. However, if we use some other algorithm to recover the space then indexing can be of value to allow rapid lookup of neighbours. To do this the tree algorithm would need to handle a growing space as the algorithm progresses, for which we are not aware of any easily applicable framework.

%Inference of this space works well when the dimension of the space is small (e.g. \cite{fang2012euclidean,alfakih1999solving}). The matrix completion literature above uses Singular Value Decomposition to understand the structure of the underlying entities.

\subsection{Active learning} \label{sec:active}

Some supervised classification problems have the following
characteristics. First, there is a small amount of labelled data and a
large amount of unlabelled data. Second, the process of labelling is
costly. In these scenarios, active learning \cite{cohn1996active} is used to select
unlabelled data for presentation to an oracle for labelling. The
objective is select those unlabelled data that will yield the greatest
improvement to the classification model, thereby minimising the cost
and maximising the utility of  labelling. There are numerous
variations on this basic theme. \cite{settles2010active} provides an excellent
review, and details the many heuristics that have been
explored. Interestingly, it seems only recently that performance
comparison against a random selection benchmark is important. 

Our framework shares many characteristics with Active Learning: we
seek to score unevaluated similarities $S_{ij}$ with the intention
that the scores are indicative of the value of the similarity to the
task. \cite{candes2010matrix} discusses the role of an oracle
in matrix completion problems. As with active learning, random
selection of similarities is an important benchmark in our framework. Unlike most work in Active Learning, we obtain a continuous outcome, we do not have an explicit feature space in which to work, and we are interested in different loss functions.

\subsection{Sequential decision making} \label{sec:sequential}

Sequential decision often involves selecting actions to maximise
reward in noisy or uncertain environments. A good overview is
given in ~\cite{cesa-bianchi2006}. The
``exploitation-exploration dilemma'' is always present in such
problems. This dilemma concerns the competing objectives of selecting
actions for maximal gain (exploitation) and for reducing uncertainty
(exploration). The core of our framework is sequential selection of
objects $S_i$ to evaluate, and the exploitation-exploration
dilemma naturally arises. 

One simple approach to address the dilemma is to incorporate an
element of randomised decision making. Such methods, called
$\varepsilon$-greedy methods, select the greedy action (the action
with highest predicted reward) with probability $1-\varepsilon$, and a
random non-greedy action with probability $\varepsilon$. Such
algorithms have solid theoretical justification for infinite horizon
problems and have been demonstrated to behave well empirically.  

\section{Methodology} \label{sec:method}
We start in Section \ref{sec:general} with the general decision framework for ordering computation. Section \ref{sec:simple} illustrates the advantage of reasoning about computation with the well known example of an autoregressive model. Section \ref{sec:dds} explains the Data Directional Scaling method for constructing a similarity space. Section \ref{sec:similarity} defines the similarity problem precisely, whilst \ref{sec:doemulation} fully defines the emulation framework. Section \ref{sec:choice} defines the choice framework, whilst the simpler issues of Assessment (Section \ref{sec:assessment} ) and Termination (Section \ref{sec:termination}).  Finally, this is wrapped up with a simulation study in Section \ref{sec:simstudy}.

\subsection{General framework}\label{sec:general}

Consider a discrete set of objects $i=1 \cdots M$ about which we can choose to take measurements (hereafter called computations) $S_{i=1 \cdots M} =\{S\}$ at cost $L_i$, with mean cost $\mathbb{E}(L_i)=L$. If $M L$ is large relative to our computational budget, we cannot afford to compute them all. Despite this, we wish to compute some quantity, say $\phi(\{S\})$, of the whole dataset.  How well we estimate $\phi$ is defined by a loss $\mathcal{L}(\{S\})$.  For example, if $\phi=\bar{S}$ is the sample mean then we could define $\mathcal{L}(\{S\}) = \var(\phi)$ to specify a minimum variance solution.

We will iteratively make choices about which object $i(t)$ to evaluate at iteration $t=1 \ldots T_{max}$. The number of iterations $T_{max}$ may not be fixed. Let $\Omega(t) = \{i(1), \ldots ,i(t)\}$ be the set of objects selected up to now, and $S^*(t) = \{S_{\Omega(t)}\}$ be the associated observed quantities. Our framework uses two key concepts:
\begin{itemize}
\item {\bf A Choice function}: $\mathcal{C}(S^*(t))$ takes the previously observed information $S^*(t)$ and chooses an object $i$ from the unobserved objects $\Omega^\perp(t)$.
\item {\bf An Emulator}: $\mathcal{E}(i;S^*(t))$ The emulator takes the previously observed information $S^*(t)$ and returns a predictive distribution $\hat{S}_i(t)$ on any observed object $i$.
\end{itemize}
The choice function $\mathcal{C}(S^*(t))$ can use any of the available information. Specifically, the choice function can use the emulator $\mathcal{E}(i;S^*(t))$ to make intelligent decisions via the following tools:
\begin{itemize}
\item {\bf A Loss estimator}: $\hat{\mathcal{L}}(S^*(t)) = \mathbb{E}(L(\hat{S}))$ over possible values of $\hat{S}_i(t)$. Note that $\hat{S}_i \equiv S_i$ for $i \in \Omega$.
\item {\bf A Heuristic}: $\mathcal{H}(S^*(t))$ is a decision rule avoiding emulation. In some special cases, it can be shown to minimise a loss $\mathcal{H}(S^*(t)) = \argmin \hat{\mathcal{L}}(S^*(t))$. If it does, it is \emph{exact} for that loss.
\end{itemize}

Although emulators that make a point prediction could be exploited, most interesting loss functions require as assessment of uncertainty.  We will explore the loss function more fully in Section \ref{sec:choice} after introducing the similarity problem.

\subsubsection{Computational constraints limit choice}

The purpose of this framework is to reduce computation. As this restricts the number of quantities that the Choice function can evaluate, it is helpful to describe it in more detail. $\mathcal{C}(S^*(t))$ proposes a number $m(S^*(t),t)$ of objects $\Omega^{(t+1)}_m$ for which the minimum predicted loss $\argmin_m \hat{\mathcal{L}}(S^*(t)| \Omega^{(t+1)}_m)$ is estimated. A special case is where there is only one proposal, in which case no loss calculation is required. This can happen if an exact heuristic $\mathcal{H}(S^*(t))$ is available for the loss $\mathcal{L}$. Even if there is no known exact heuristic, proposals from other heuristics are often helpful.  

Finally, note that the Choice function has available the current iteration $t$ as well as being able to assess the success of previous choices. This can be used choose to enable or disable proposals. For example, it is often useful to use more complex proposals initially, and then switch to simpler proposals as the marginal return of careful choice decreases.

\subsubsection{Full choice algorithm}
The algorithm proceeds as follows. For each iteration $t$:
\begin{enumerate}
\item {\em Choice}: 
  \begin{enumerate}
  \item {\em Proposal}: $\mathcal{C}(S^*(t))$  proposes objects $\Omega_m^{(t+1)}$ to evaluate.
  \item {\em Emulation}: If $M >1$, estimate the loss $\hat{\mathcal{L}}_m(\hat{S}(t+1) | \Omega_m^{(t+1)})$. 
  \item {\em Decision}: Choose the action $m^\prime$ that minimises the loss.
  \end{enumerate}
\item {\em Evaluation and Assessment}: Compute $S^*(t+1)$ of the chosen objects $\Omega_{m^\prime}^{(t+1)}$. Compare the predictive distribution $\hat{S}_{\Omega_{m^\prime}^{(t+1)}}(t)$ to the observations $S^*_{\Omega_{m^\prime}^{(t+1)}}$.
\item {\em Termination}: Stop if a stopping criterion is reached.
\end{enumerate}
These processes are precisely defined below.  Specifically, Section \ref{sec:doemulation} deals with emulation, which also uses the results of assessment. Section \ref{sec:choice} deals with the remaining aspects of the Choice function. Section \ref{sec:termination} addresses stopping the algorithm. 

\subsubsection{Simple example} \label{sec:simple}
Before we narrow our attention to similarity matrices, it is helpful to demonstrate the value of reasoning about computation more generally. As an example, let $S_i$ (with $i=1 \cdots M$) describe an autoregressive (AR1) model:
$$
S_i = \psi + \phi S_{i-1} + \epsilon_i
$$
with $\phi<1$ and $\epsilon_i \sim N(0,\sigma_\epsilon^2)$. Let us assume that the process has reached stationarity. We wish to estimate $\mathbb{E}(S) = \mu = \psi/(1-\phi)$ via an estimator $\hat{\mu}$. If either $M$ is large, or the cost per computation $L$ limits the number we can make, then we cannot compute $\mathbb{E}(S)$ using $\bar{S} = \sum_{i=1}^M S_i$. However, we can obtain a Monte-Carlo estimate by evaluating a subset $T_{max} < M$.  Since they are correlated, which should we choose?

Let $\hat{\mu} = \sum_{i \in \Omega(T_{max})} S_i$, and the loss be $\mathcal{L}(S) = \var(\hat{\mu})$. From the standard Monte-Carlo methodology (e.g. \cite{Gamerman1997}) it is known that minimising this variance is the same as minimising the covariance between the samples. This can be performed explicitly to result in an optimal heuristic.

If $T_{max}$ is known in advance then $\mathcal{H}(S^*(0))$ is known, i.e. we can choose $\Omega(T_{max})$ before we compute any $S$. This uses evenly spaced samples with separation $N/T_{max}$ (called thinning in the Markov-Chain Monte Carlo literature).  If however $T_{max}$ is not known in advance, $\mathcal{H}(S^*(t))$ can use an iterative greedy approach by picking the object $i(t+1)$ furthest from all previous objects $\Omega(t)$, i.e. select objects $1,N,N/2,N/4,3N/4$ etc.  If the autocorrelation length of $S$ is greater than $T_{max}$ then this approach massively reduces the variance over the naive approach of using the first $T_{max}$ objects. It also clearly improves on the selection of random objects, which is a heuristic that is frequently helpful when nothing else can be done.

If our desired loss function is to instead estimate the parameters $\psi$ and $\phi$ as accurately as possible, then we must decide on an estimation framework (least squares, Yule-Walker equations, maximum likelihood via numerical optimization, etc. \cite{hamilton1994time}). Then we face a tradeoff between obtaining high lags to estimate $\mu$ and low lags to estimate $\phi$. In this case, the correct model can be used via a Kalman filter \cite{kalman1960new}. This can be seen as Emulation, and is one way to handle missing data \cite{howell2007treatment,jones1980maximum}. Developing the appropriate loss and choice functions is possible but outside the scope of this paper.

\subsection{Data Directional Scaling} \label{sec:dds}

Before we go into the specifics of decision making for similarity matrices, we introduce `Data Directional Scaling' (DDS) which is a type of Multidimensional Scaling (MDS \cite{Young87}) that has proven very helpful for our emulator. In MDS, the data is seen as having a position on a manifold or metric space described by a basis. There are a wide range of MDS methods, but the most commonly used is principal component analysis (PCA or equivalently, singular value decomposition, SVD). In PCA, basis vectors are chosen to be orthogonal and decreasing in their contribution to the variance observed in the dataset.

For the similarity problem, consider a set of objects $i=1 \ldots N$ describing data  $D_{il}$ with $l=1 \cdots L_{i}$ indexing information about those objects. There is not necessarily a consistent dimension to the data, but a non-negative similarity function $S_{ij} = S(\mathbf{D}_{i},\mathbf{D}_{j})$ can always be evaluated. If additionally $S$ forms a metric space then $S_{ii}=S_0$ for all $i$ and we can write distances $X_{ij} = S_0 - S_{ij}$. $X_{ij}$ are also non-negative with $X_{ij}=0$ if and only if objects $i$ and $j$ are identical under $S$. Conversely, if $S$ is non-metric, we can still convert to a distance-like measure via $X_{ij} = \max_{k}S_{kk} - S_{ij}$ but some self-distances are non-zero. We can interchangeably work with $S$ or $X$ as we do not exploit any properties of a metric.

Without access to any space in which the objects can be embedded, most traditional inference frameworks are difficult to apply.  The matrix completion approach seeks to find the underlying space by inferring a low-rank latent space (typically an SVD). However, this fails when insufficient matrix elements have been evaluated.

\begin{figure} [ht]
\begin{centering}
\includegraphics[clip,width=0.45\textwidth,angle=0]{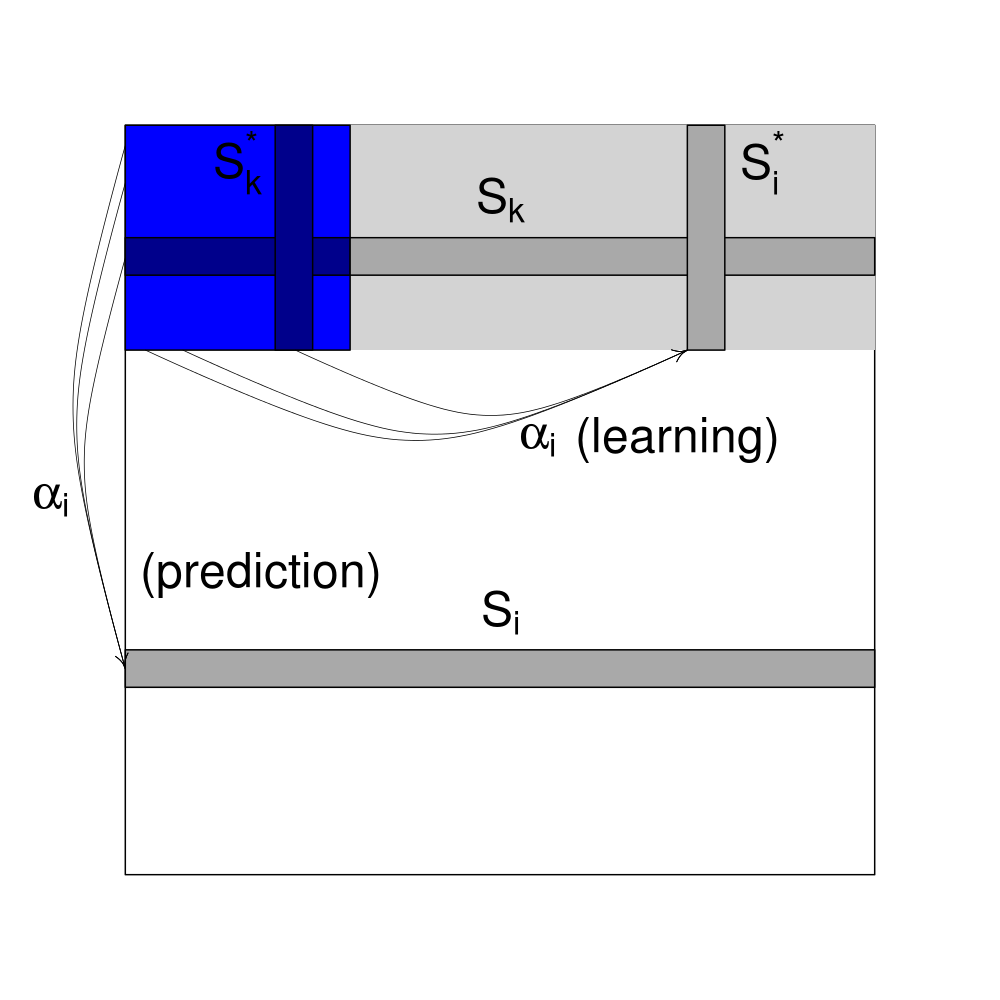}
\includegraphics[clip,width=0.45\textwidth,angle=90]{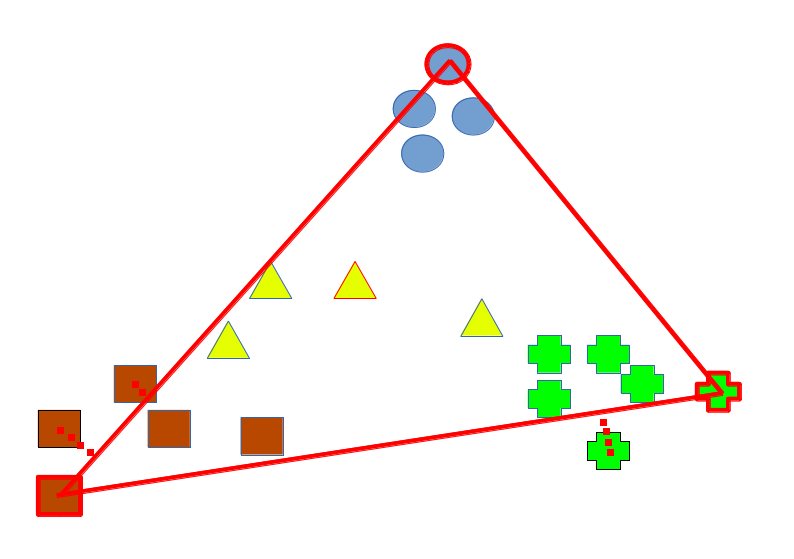}

\caption[] {Data Directional Scaling illustration.  Left: Quantities used in DDS. Observed rows of $S$ are shown at the top, with the vector space $\mathcal{S}(t)$ being defined by the set of columns of observed similarities $S^*_{k}$ for observed objects $k \in \Omega{t}$. Object $i$ is associated with observations $S^*_i$, which is mapped to a position $\alpha_i$ in the vector space using the learning model for the Emulator $\mathcal{E}(i;S^*(t))$.  The prediction space is the rows $S_i$ of the matrix, for which we observe $S_k$ with $k \in \Omega(t)$.  Right: Example with an underlying Euclidean space $\mathbb{R}^2$. If we have observed the three objects in red, then we can reconstruct any point in the $2D$ simplex (i.e. triangle) shown in red. Positions outside the simplex are mapped to the nearest point inside the simplex.
\label{fig:dds} }
\end{centering}
\end{figure}

Our approach evaluates matrix elements in rows $\mathbf{S}_{i}$ by calculating all similarities with object $i$.  This is called `evaluating object $i$'. It leads directly to a vector space $\mathcal{S}(t)$ using only the information from observing objects $\Omega(t)$. The basis of $\mathcal{S}(t)$ is the \emph{columns} $\mathbf{S}^*_b$ of the similarity of each evaluated object $k \in \Omega(t)$ with each other object $b \in \Omega(t)$.  Therefore an additional observation at iteration $t+1$ simply creates a new direction in the space.  Figure \ref{fig:dds}a illustrates this procedure.

This is useful because columns $\mathbf{S}^*_i$ for $i \in \Omega^\perp(t)$ are observed. The emulator $\mathcal{E}(i,S^*(t)$ provides a position $\alpha_i(t)$ for object $i$ in the space $\mathcal{S}(t)$. For practical reasons, we define $\alpha_i(t)$ to live in the $t-1$ dimensional simplex; i.e. $\sum_{\alpha_i}=1$ and $\alpha_i>0$ for all $i$.

Now our objects $i$ are standard observations in a vector space, and we can use standard tools to perform inference. Specifically, we treat columns $S^*_i(t)$ as observations, and rows $S_i(t)$ as responses. We can therefore use standard linear methods to predict unobserved rows $S_i(t)$. This procedure is discussed in Section \ref{sec:doemulation}.

Figure \ref{fig:dds}b interprets this procedure when objects are truly embedded in $\mathbb{R}^2$, which is a metric space and therefore the similarities are symmetric.  If we further assume that similarities are observed with noise, then triangulation is non-trivial. By observing objects at the `corners' of the space, the unobserved objects fall inside or near the convex hull of the observed objects. In this case, we will be able to accurately represent the whole space without extrapolation.

\subsection{Similarity model} \label{sec:similarity}

We will use DDS and the choice framework to optimally select elements from the similarity matrix to evaluate.  The expected cost of a similarity computation is defined to be $L$, with $L \propto \mathbb{E}(L_i)$ if the computation is linear. The total cost of computing this matrix is therefore $O(N^2 L)$, which we assume is too large. Our methodology will reduce this computation to $O(N^2 + NL)$ for quite general problems. This can be reduced to $O(NL)$ if a) the full matrix is not required and b) either a less exact solution is satisfactory, or explicit solutions to certain sub-problems (i.e. exact heuristics) can be found.

%We split the computation into a number of repeated timesteps $t=1 \cdots T_{max}$, at which we evaluate one or more entries of $S$.  Let $\mathbf{h}_t$ be the history of choices made before time $t$ and $C_t$ be the matrix representation of the evaluations chosen at time $t$, taking value 1 for chosen pairs and 0 otherwise.  Further, $C^t = \sum_{t^\prime=1}^tC_{t^\prime}$ is the induced matrix elements chosen up to and including time $t$, $S^*(t)$ is the observed matrix, excluding missing values, and $S(t) = S(t ;\mathbf{h}_t)$ is the observed matrix at that time (with missing values), and $\hat{S}(t)$ is the emulated values.  At time $t=0$ the matrix is unobserved, $C^0=C_0=0_{N \times N}$.  Due to the way our calculation progresses, we will frequently evaluate $S$ in rows; the index of the rows evaluated up to time $t$ is $\Omega(t)$.

As above, we use a choice function and an emulator. The Choice function  $\mathcal{C}(S^*(t))$ now can return either a single similarity element (i.e. an $i,j$ pair), or a set of elements. To use DDS, it is necessary to evaluate $S_{ij}$ in entire rows.  The emulator still returns the value of single elements, and takes the form 
\begin{equation}
\mathcal{E}(i,j;S(t)) = \hat{S}_{ij}(t) \sim N(\bar{S}_{ij}(t),\sigma_{ij}^2(t)) \label{eqn:emulator}
\end{equation}
which provides a distribution for any element $i,j$. $\bar{S}_{ij}(t)$ and $\sigma_{ij}^2(t)$ both require further specification.  Whilst it is arguable whether the Normal model is appropriate, we note that it is justifiable on computational grounds. Further, if the similarities are sums over features then (as in Gaussian Process emulation) the Central Limit Theorem justifies this choice.

\subsection{Emulation} \label{sec:doemulation}
Recall that in our framework, the utility of an emulator is to be able to predict how calculating a particular set of elements will improve our ability to optimise a loss. We therefore will need to be able to compute many emulated values, and recompute our emulator as new information arises. A slightly separate use is to `complete' the matrix at the end of the algorithm to provide a `best estimate'. Compare this use with matrix completion, for which the computational budget is allowed to be larger.

For defining the emulator, we abuse our notation and set $\Omega(t)$ to be the index of the objects we have evaluated, i.e. \emph{full rows} of the matrix $S$ evaluated up to time $t$.  The emulator takes the form:
\begin{equation}
\hat{S}_{ij} \sim N\left( \bar{S}_{ij}(\Omega(t)), \delta_{i}^2(\Omega(t))\right), \label{eqn:emulation}
\end{equation}
where we have dropped the obvious dependence on $S^*$. The quantities $\delta_i$ and $\bar{S}_{ij}$ are both inferred using linear prediction frameworks, due to computational restrictions.

\subsubsection{Prediction of the mean}
Using the Data Directional Scaling from Section \ref{sec:dds} and Figure \ref{fig:dds}a, $\bar{S}_{ij}(\Omega(t))$ is estimated as follows. For $k \in \Omega(t)$, a linear regression is used to \emph{learn} the observations in column $i$ using the observed columns $k$:
$$
S^*_{ji, j \in \Omega(t)} = \sum_{k \in \Omega(t)} \alpha_{ik} S^*_{jk} + e_{ji}(t).
$$
Row $i$ is \emph{predicted} using the observed rows:
$$
\bar{S}_{ij, j \in 1 \cdots N} = \sum_{k \in \Omega(t)} \alpha_{ik} S_{ij} + d_{ij}(t).
$$
Here, $e_{ji}(t) \sim N(0,\epsilon_i^2(t))$ is the standard residual error on observed elements $S^*(t)$. $d_{ij}(t) \sim N(0,\delta_i^2(t))$ is the out-of-sample prediction error for an unobserved element $S_{ij}$. The residual $\epsilon_i^2(t)$ is useful, but we will are particularly interested in the understanding the `emulator variance' $\delta_i^2(t)$, the same quantity from Equation \ref{eqn:emulation}.

As discussed in the DDS section, $\alpha_i$ is learnt under the constraint that $\sum_{k} \alpha_{ik} = 1$ and all $\alpha_{ik} \ge 0$ using one of the following estimation procedures:
\begin{itemize}
\item[NN:] Nearest Neighbour. Choose $\alpha_i = \argmin_\alpha(\| \sum_{k \in \Omega(t)} \alpha_{ik} S_{jk} -  S_{j i}\|_\infty)$. By weighting large similarities `infinitely' strongly this norm leads to selecting $\alpha_{ik} =1$ for $k=\argmin_j(S_{ji})$, i.e. the nearest neighbour in the column $S_{ji}$ with $j \in \Omega(t)$, and zero otherwise. 
\item[MM:] Mixture Model. Choose $\alpha_i = \argmin_\alpha(\| \sum_{k \in \Omega(t)} \alpha_{ik} S_{jk} -  S_{j i}\|\|_2)$, i.e. fit the mixture to minimise the square error. This is inferred using quadratic programming \cite{quadprogR}.
\end{itemize}
Figure \ref{fig:dds}b captures the mixture model. We also considered an unconstrained linear regression, but this is unstable because the inference is overspecified when $t > R$; in practice two objects within one cluster are used with a very large positive and negative weight.

Because we are dealing with similarities, it is necessary to handle `self' specially. We swap $S_{kk}$ and $S_{ki}$ so that the estimator `sees' the similarity of $k$ with $i$ instead of the that of $k$ to $k$ (which is only zero for metric spaces).

\subsubsection{Prediction of the Emulator Variance}
Estimating $\delta_i$ must be rapid, and capture the distribution of the sampled objects and structure of the manifold on which they lie. In principle it is a function of the full set of similarities with the evaluated objects, and is not `observable' in the way $\bar{S}_{ij}$ was, because the uncertainty of an observed element is zero. We calculate the uncertainty of the prediction via a linear combination of most conceivable summary statistics:
\begin{equation}
\delta_{i}^{\mathrm{full}}(\Omega(t)) = \gamma_0 + \gamma_t t  + \gamma_{t^{-1}} t^{-1} + \gamma_\epsilon \epsilon_i(t) + \sum_{n \in \mathcal{N}} \gamma_{n} \bar{R}_n(i;\Omega(t))+\mathrm{residuals}. \label{eqn:deltafull}
\end{equation}
This includes coefficients for an intercept $\gamma_0$, the number of previously evaluated objects $t$ via ($t$ and $t^{-1}$), and the regression residuals $\epsilon_i(t)$.  We also account for the full similarities distribution of  object $i$ with evaluated objects $k \in \Omega(t)$ using `similarity distance' in the $L_n$ norm (using $n \in \mathcal{N}=\{1,2,\infty\}$): 
$$\bar{R}_n(i;\Omega(t)) = \left(\sum_{i \in \Omega(t)}\|S_{ji}\|^p \right)^{1/n}.$$

{\em Fitting the emulator variance $\delta_i(t)$}: The set of objects $\Omega(t)$ chosen for evaluation may be very far from random and $S_{jk}$ (for $j,k \in \Omega(t)$) are typically biased to small values (i.e. the distances between them are larger than average, as in Figure \ref{fig:dds}b). Extrapolation using linear regression can therefore be misleading and we enforce a sensible prediction by using non-negative least squares regression (NNLS,\cite{lawson1974solving}) instead. 
This automatically sets some coefficients to zero, although explicit penalisation can also be used. 

Empirically, our NNLS metho most commonly chooses the regression:
\begin{equation}
\delta_i(\Omega(t)) = \gamma_\epsilon \epsilon_i(t) + \gamma_{\infty} \bar{R}_\infty(i;\Omega(t))+\mathrm{residuals}. \label{eqn:deltaused}
\end{equation}
This is computationally efficient to work with as only nearest neighbours have a changed $\gamma_{\infty}$ coefficient and $\epsilon_i(\Omega(t))$ changes only by a single coordinate, leading to an efficient calculation for $\delta_i(t+1)| \delta_i(t)$. Therefore the results we report use Equation \ref{eqn:deltaused}, although we have implemented Equation \ref{eqn:deltafull} and checked that its performance is not significantly different.

{\em Cross validation estimate of $\delta_i(t)$}: We have a choice for how we fit $\delta_i(t)$ to achieve the best predictive power for the lowest computational effort. We have available to us, for free, the observed values of $\delta_j(\Omega(t^\prime))$ for the previously chosen objects $j \in \Omega(t)$ for previous times $t^\prime\le t$ with a different set of observed objects  $\Omega(t^\prime)$.  This acts as a `poor mans cross validation'. We can either:
\begin{enumerate}
\item Use only the observed history;
\item Use only cross-validated estimates from the current observations $t$, by treating each object $i \in \Omega(t)$ as if it were the last object to be evaluated;
\item Combine the approaches by retaining any cross-validation performed at previous iterations and combining it appropriately with cross-validation at iteration $t$.
\end{enumerate}
Using only the observed history will cause problems when there are few evaluated objects. Additionally we might expect `non-stationarity' in that early objects are measured in a different space and hence follow a rather different distribution to later objects. Conversely, using only cross-validation at the current iteration is computationally costly, does not account for any time-dependent learning, and exposes only a limited set of distances. Combining the approaches would seem to be appropriate but requires choosing a way to `age-off' less accurate information from early in the process. Additionally, we might wish to choose how much cross-validation to perform; intuitively, less is needed later in the process. 

We therefore define:
\begin{itemize}
\item {\bf A cross-validation operator}  $\mathcal{V}(t)$ which decides which historical objects to retain.
\end{itemize}
By default we use $\mathcal{V}_{obs}(t)$ which uses all the history up to time $t$ (excluding the first 5). 

We have an additional choice about how to handle the additional information about $S_{ik}$ when $k \in \Omega$ if $\mathcal{S}$ is known to be symmetric.  First, we can set $\hat{S}_{ik}=S_{ki}$ and use the above framework. Second, we can use its inferred value to calibrate $\delta_i(t)$, estimating $\delta_i(t) = \sum_{k \in \Omega(t)} (\hat{S}_{ik} - S_{ki})/t$ instead of the regression described above. We have not investigated these options in detail, instead treating all similarities as non-symmetric.

\subsection{Choice: Proposals and Decisions} \label{sec:choice}
We can exploit the emulator as defined to make intelligent decisions about which objects and/or matrix elements to evaluate using the choice function $\mathcal{C}$. The loss depends on the emulated matrix $\hat{S}(t)$ via the evaluated similarity elements $S^*(t)$. It also depends on the computation spent. Let $L_{ij}$ be the computational cost for computing element $S_{ij}$. Then for simplicity we consider losses of the form
$$
\mathcal{L}(S^*)  =\mathcal{L}_0(S^*(t)) + \mathcal{L}_1\left(\sum_{i,j \in \Omega(t)}L_{ij}\right),
$$
where $\mathcal{L}_0$ is the `model fit' loss, and $\mathcal{L}_1$ is the `computational cost' loss. This simplification means that we can minimise each loss separately and easily combine them. If we further assume that $L_{ij}=L$ for all pairs, then the computational cost of any proposal containing the same number of elements is the same. For notational simplicity we therefore omit the subscript $\mathcal{L}_0$ below.

\emph{Loss for model fit}: A fairly general loss is a (weighted) mean prediction error of the form:
\begin{equation}
\mathcal{L} = \mathbb{E} \left(\sum_{ij} w_{ij}(\hat{S}_{i j}) \|\hat{S}_{i j}- S_{ij}\|_n \right), \label{eqn:loss}
\end{equation}
where the weights $w_{ij}(\hat{S}_{i j})$ allow us to specify which elements are most important for a particular problem, and can allow for covariance. We consider the $L_n$ norm of the difference between the true and observed values of the matrix.  

We do not have access to the true value of $S_{ij}$ at any iteration, but the emulator provides a calibrated way to estimate the expectation for decisions.  Therefore, although we do not have access to the absolute value of the loss, we can estimate the difference in loss between decisions with high accuracy, and hence make good decisions.

It is helpful to distinguish between two classes of evaluation that we use:
\begin{itemize}
\item {\bf Global search:} To identify bulk cluster structure, propose to evaluate an entire row of $S$, setting $\Omega_m^{(t+1)} =i$. Evaluate all similarities $S_{ij}$ of an object $i$ with all other objects $j$.
\item {\bf Local search:} To identify local neighbourhood structure, propose to evaluate a single element $i,j$ with value $S_{ij}$. This is given the notation $\Omega_m^{t}=i,j$, and if chosen, $\Omega(t)$ now includes this element. i.e. $t$ is \emph{not} incremented.
\end{itemize}

\subsubsection{Global Search} 
If we were to evaluate all similarities with object $i$, then $\Omega_m^{(t+1)} =i$. This has two effects on the loss. First, it evaluates $\hat{S}_{ij}$ to be $S_{ij}$, and second, it can be used to improve predictions of other objects via the emulator. Evaluating the improvement of the loss in principle requires integration over all assignments of $\hat{S}_{ij}$.

However, we can obtain a good estimate of $\hat{\mathcal{L}}_m(\hat{S}(t+1) | \Omega_m^{(t+1)})$ for significantly less compute by `plugging-in' the expected value $\bar{S}_{ij}$.
We therefore assume that a given row will be the mean predicted value, and predict the change in loss for the rest of the matrix.  Therefore we can write the loss and therefore the optimum choice $\mathcal{C}(t+1)$ as
\begin{eqnarray}
\hat{\mathcal{L}}_m(\hat{S}(t)) &=& \sum_{ij} w_{ij} (\bar{S}_{i j}) \|\delta_i(t)\|_n \\
\mathcal{C}(t+1) &=& \argmin_{m} \left(\hat{\mathcal{L}}(\hat{S}(t)) - \hat{\mathcal{L}}_m(\hat{S}(t+1| \Omega_m^{(t+1)}) \right) \label{eqn:lossprediction}
\end{eqnarray}
where $\delta_i(t)$ for observed rows is defined to be zero. Notice that this loss depends directly on the emulator variance only. It depends on the emulator predictions via their interaction with the variance (and potentially via the weights). The expectation in Equation \ref{eqn:lossprediction} is easily evaluated by estimating $\delta_j(t+1|\Omega_m^{(t+1)})$ via the same regression from Equation \ref{eqn:deltaused}, with the set of objects $\Omega_m^{(t+1)}$.

We define one loss and two heuristics via this framework.
\begin{itemize}
\item \emph{RMSE loss}: The $l_2$ norm is a natural measure and can be used to directly minimise the root-mean-square error (RMSE) in the prediction. This is a matrix completion approach. This loss can be computed by brute force in $O(N^2)$ by evaluating the effect of each observation $i$ on the corresponding variance $\delta_j^2$.
\item \emph{Furthest distance heuristic}: If we use uniform weights and the $l_{\infty}$ norm for the loss, then the largest value of $\delta_i$ dominates. The loss is minimized by evaluating the object that is furthest away from all evaluated objects. This heuristic is inspired by the nature of the space from Section \ref{sec:dds}.
%This `furthest point' heuristic is an important building block, because it can be evaluated in better than $O(N)$ time. First the minimum distance for each column is calculated, and then the maximum distance is sought. By caching previous locations, the amortised cost can be $O(\log(N)^2)$ by constructing an appropriate tree (e.g. an R tree), which happens once the observed points span the underlying manifold. 
%\item \emph{Highest variance heuristic}:  If we use place weight only on $i$ and use the $l_{2}$ norm for the loss, then the loss is minimized by evaluating object with the largest emulator variance.
\item \emph{Random selection heuristic}: Whilst not strictly available under the loss framework, if all weights are zero the loss is uniform and we can select at random. This is an important benchmark.
\end{itemize}

In practice, evaluating the full emulated RMSE loss function is too costly. It is useful instead to use a Monte-Carlo estimate of it. We estimate the loss using a constant number $p=\min(80,N-t)$ of random unevaluated objects (with the addition of one chosen by the furthest distance heuristic). We then estimate the loss using those samples only, reducing the computation from $O(N^2)$ to $O(p^2)$. 

\subsubsection{Local Search} 
% We use a graph-based algorithm described below based on nearest neighbours to identify close points? OR: We evaluate within a cluster? Do we have an online method of finding clusters? Divisive hierarchical clustering? treating everything as multivariate normal?
Local search permits evaluation of elements without calculating an entire row. However, our emulation framework cannot exploit this information without additional complication. We therefore allow local search to only influence the loss at specific values of $i,j$ and omit any learning that this point provides about the rest of the matrix. For some loss functions, however, it is extremely important to calculate certain matrix elements at the omission of others and therefore these choices might be applicable.  We can choose local proposals using the same loss function as in the global case, and indeed decide between choices on the basis of this loss. 

In this case we need to consider $\mathcal{L}_1$ in the loss. In practice we have used a linear function $\mathcal{L}_1(x)=c x$. In this case the cost of a local proposal is $cL$ and the cost of a global proposal is $cLN$. The factor $c$ is chosen to convert computational time units to `loss per iteration' units.

\subsection{Evaluation and Assessment}\label{sec:assessment}
Evaluation and assessment are straightforward. The matrix elements are evaluated, and compared to the predictive distribution. The predicted mean of the emulator $\bar{S}_{ij}(t)$ is compared to the observed values $S_{ij}$ using its RMSE:
$$
\mathrm{RMSE}(\bar{S}_{ij}(t)) = \sqrt{\frac{1}{N}\sum_{j=1}^N \left(\bar{S}_{ij}(t) - S_{ij}\right)^2}.
$$
The predicted emulator variance $\delta_i^2(t)$ of the emulator is also compared to the empirical residual variance $\delta^2_{obs,i} = (1/N)\sum_{j=1}^N \left(\bar{S}_{ij}(t) - S_{ij} \right)^2$ using its error:
$$
\mathrm{Error}(\delta_{i}(t)) = (\delta_i(t) - \delta_{obs,i}(t))^2.
$$
This is much noisier since we only obtain one error per object observed.

In both cases, it is of value to determine convergence which could be used in the termination criteria. There are standard test statistics that could be used.  We have not tried to terminate the algorithm on this basis because the loss functions we have used do not converge, but instead slow down. Other loss functions may converge, however.

\subsection{Termination} \label{sec:termination}
There are three fundamental ways of running the algorithm, which can be defined using an appropriate computational loss term $\mathcal{L}_1$. Termination can be seen as an option that minimises the loss $\mathcal{L}$, when all possible actions result in a decrease in loss. 
\begin{enumerate}
\item Obtain the minimal loss for a fixed computation. We have seen in Section \ref{sec:simple} that it is helpful to think about the total number of evaluations that will be available. However, most calculations are not this structured and we can run our algorithm until a computational budget is reached. The corresponding computational loss $\mathcal{L}_1$ takes an arbitrary function up to the budget $B$ when it is set to $\infty$.
\item Obtain the cheapest estimate with a fixed precision. If a desired precision of the loss is known \emph{apriori}, then we can use standard results from sequential stopping rule theory \cite{lindley1961dynamic,berger1985statistical} under the assumption that the loss has normally distributed error. Again, $\mathcal{L}_1$ takes a step to $\infty$ but now when the precision is reached.
\item Use the full loss $\mathcal{L}_1$. We would then be able to explore the cost/benefit ratio of obtaining further similarities by making further computations. 
\end{enumerate}
We have here only worked with criterion 1, i.e. we compare the performance for a given amount of compute. Further, in the simulated example, the cost of each computation $L$ is arbitrary and so we do not compute the real computation performed but estimate it to a factor of $L$. 

\subsection{Simulation Example}\label{sec:simstudy}
\begin{figure} [ht]
\begin{centering}
\includegraphics[clip,width=1.0\textwidth,angle=0]{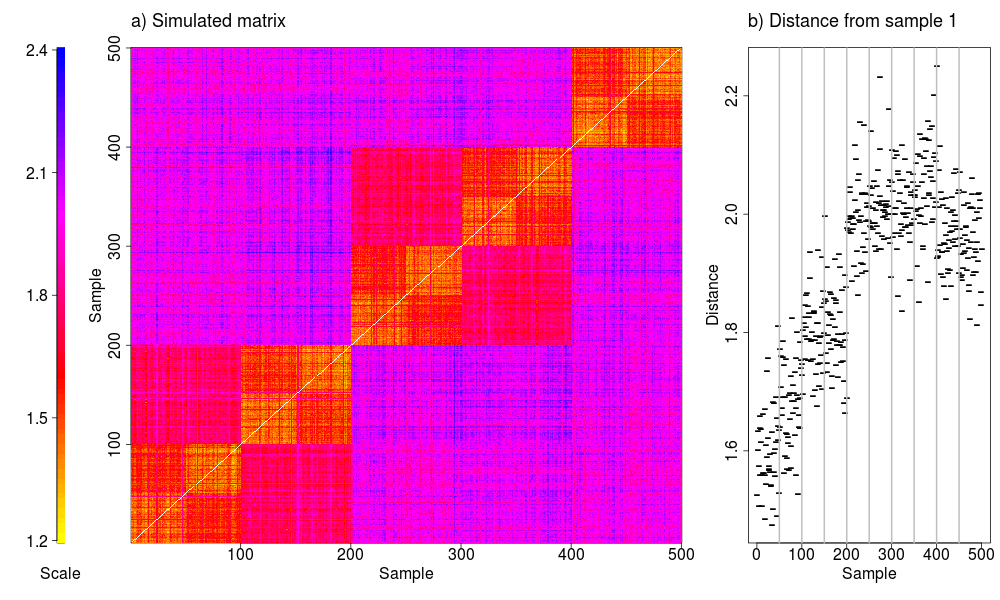}
\caption[] {Simulated dataset, showing a) the whole `corrupted' matrix as a heatmap, and b) the distribution of distances from an arbitrary row. Note that neighbouring clusters overlap strongly, and next-nearest clusters have overlapping distributions. Outliers appear as lower intensity rows and columns, and clusters contain 50 samples.
\label{fig:SimData} }
\end{centering}
\end{figure}

We illustrate our methodology with application to a simulated dataset. $L=1000$ Gaussian features were simulated by assuming a tree correlation structure for $N=500$ samples in $R=10$ evenly sized clusters. This type of dataset is representative of many real world applications, including genetics. It is also challenging for standard matrix completion approaches because the eigenvalues correspond to the clusters \cite{lawson2012population}, which might not be informative about all samples. This assumption is required for matrix completion to have performance guarantees \cite{candes2010matrix}.  Figure \ref{fig:SimData} illustrates one such matrix.

To assess the robustness and performance characteristics of the algorithm, we vary the difficulty of the problem. For this we vary the correlation within and between clusters, and allow outliers of varying sizes.  Table \ref{tab:simpar} describes the parameters used for the datasets we generated, which lie on a continuum from `clustered' in which inference is simple, to `corrupted' which contains many outliers and has weaker correlation structure. 

\begin{table}[h]
\centering
\begin{tabular}{c|l|c|c}
Parameter & Description & Start `clustered' & End `corrupted' \\
\hline
$c_0$ & Within cluster correlation & 0.75 & 0.3 \\
$c_1$ & Close cluster correlation & 0.5 & 0.25 \\
$c_2$ & Distant cluster correlation & 0.25 & 0.15 \\
$a_h$ & Inverse Outlier $\alpha$ parameter & 0 & 0.2 \\
$b_h$ & Inverse Outlier $\beta$ parameter & 0 & 5 
\end{tabular}
\caption[Simulation parameters]{Simulation dataset parameters for the correlation between samples. There are 10 clusters, with close clusters being (1-2,3-4,5-6,7-8,9-10), and distant clusters are (1-4,5-8). Outliers are generated by mixing rows with weight $1-w_i$ with a `self' direction (1 on the diagonal, 0 off it). The weights $w_i = \mathrm{beta}(\alpha=1/a_h,\beta=1/b_h)$ are biased to $1$ and are exactly $1$ when $a_h=b_h=0$. 11 datasets are generated, evenly space from the `clustered' to `corrupted' parameters (referred to as a difficulty scale). \label{tab:simpar}} 
\end{table}

The `clustered' dataset contains distinct clusters, whilst in the `corrupted' dataset, clusters overlap and most samples are subject to significant deviations from their cluster.  The true rank of the easy dataset is $R=10$ (excluding correlations in the noise). The presence of outliers make the true rank $N$, although the effective rank (measured by its eigenvalue spectrum) is much closer to $R$.  Figure \ref{fig:SimData} illustrates the `corrupted' dataset.

\subsubsection{Emulator performance}

\begin{figure} [ht]
\begin{centering}
\includegraphics[clip,width=0.9\textwidth,angle=0]{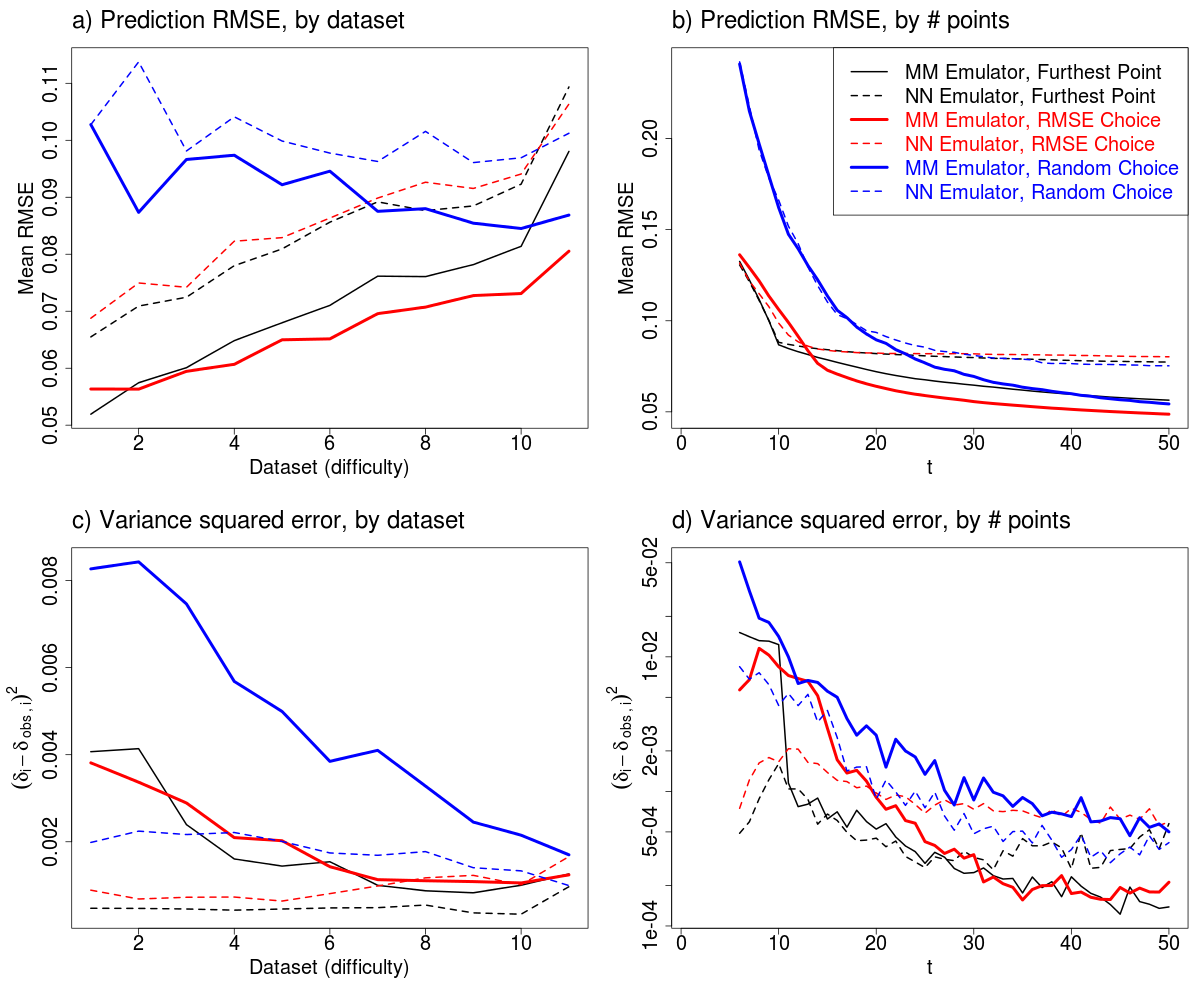}
\caption[] {Mean emulation performance averaged over 20 simulated datasets. Left: performance as a function of the complexity of the dataset, averaged over iteration number (from $1$ to $50$), from 1=`clustered' to 11=`corrupted'. Right: Performance as a function $t$, the number of objects evaluated. a-b) shows the root-mean-square prediction error (RMSE) for the unobserved portion of the matrix. c-d) shows the emulator error when predicting the `uncertainty' $\delta_i(t)$ in the estimate of the $t$-th iteration. Shown are both the Nearest-Neighbour and Mixture Model prediction methods, with objects selected using either the furthest distance heuristic, RMSE loss estimation, or random choice heuristic. 
\label{fig:PredictivePower} }
\end{centering}
\end{figure}

Figure \ref{fig:PredictivePower} describes performance (as assessed by the measures from Section \ref{sec:assessment}) over 20 replicates of the simulated datasets. Results are shown for the Mixture Model and Nearest Neighbour Emulators, choosing objects via the RMSE Loss, Furthest Distance Heuristic and Random Choice Heuristic. These are examined as a function of the number of iterations $t$ and as a function of the dataset difficulty.

The emulator performance story is consistent across all choice functions.  The difficulty of the problem is an important factor.  Figure \ref{fig:PredictivePower}a-b shows that the simpler problems with easier cluster structure demonstrate the largest difference between methods. Nearest-neighbour emulation works well for small $t$, whilst the space is still being explored, but quickly asymptotes as the number of clusters is reached. The Mixture model performs well overall and has very low asymptotic variance for this problem.   Notably, direct matrix completion approach performs significantly worse than any approach we considered. The algorithm of \cite{ma2011fixed} achieved an RMSE $=0.16$ using $25000$ random matrix elements corresponding to $t=50$. For this reason it is not considered further, although in the discussion we explain that the standard matrix completion problem is slightly misspecified for our problem.

The impact of the decision criterion on the whole matrix completion is dramatic as shown in Figure \ref{fig:PredictivePower}. For clustered datasets and early in the process, the furthest distance criterion performs very well. This success comes from first selecting objects from different clusters, but also selects outliers. For large numbers of iterations, this can be worse than choosing at random. The RMSE loss procedure, as might be expected, has the best overall performance and is robust to outliers and corrupted data. Similarly, the emulator variance estimate $\delta_i(t)$ depend on both the Choice function and emulator. 

Figure \ref{fig:PredictivePower}c-d shows how well the emulator variance is captured. Unlike the predictive performance, this asymptotes to a finite, but small, value.

Throughout, using the RMSE choice criterion and hence the Emulator dominates Random Selection Heuristic (shown in thicker lines). It also dominates the Furthest Distance Heuristic except during the early phase ($t<12$). 

\section{Computational scaling} \label{sec:scaling}

We are interested in the rate of scaling up to logarithmic terms, which in practice are negligible compared to scaling constants. Fundamentally our approach allows for linear scaling in $N$, provided that a) the matrix is of sufficiently low rank, and b) correct decision making for which elements to select requires only $O(1)$ calculation, compared to $N$. This is true for the Random Selection Heuristic and the Furthest Distance Heuristic.  The methods using matrix completion, i.e. the maximum variance criterion and the root-mean-square-error criterion, scale significantly less well, though we restricted to $O(N)$ computation by estimating the loss on a $p \times p$ matrix.

The emulator used also has an important impact on running time. Calculating the nearest neighbour of an object to a set of $t$ objects is an $O(t)$ operation, whereas inferring a mixture model using quadratic programming is only guaranteed to be polynomial \cite{monteiro1990polynomial} in $t$. In practice it is $O(t^2)$ and hence of practical value.  Therefore we suggest using this algorithm at the start of a computation, when making good decisions can strongly influence the decision making process, and is relatively efficient.

\begin{table}[h]
\centering
\begin{tabular}{l|r}
Algorithm & Order \\
\hline
Random Selection Heuristic & $N T_{max}$ \\
Furthest Distance Heuristic & $N T_{max}$ \\
%Mixture Model, Variance criterion & $N^2 T_{max}^2$ \\
Mixture Model, RMSE Loss & $N T_{max}^2 p^2 $ \\
Nearest Neighbour, RMSE Loss & $N T_{max} p^2 $
\end{tabular}
\caption[Algorithmic Scaling]{Comparison of algorithmic scaling as a function of the number of objects $N$, the number to be evaluated $T_{max}$ and the number $p$ to be used in estimating the RMSE. \label{tab:scaling}} 
\end{table}

\begin{figure} [ht]
\begin{centering}
\includegraphics[clip,width=1.0\textwidth,angle=0]{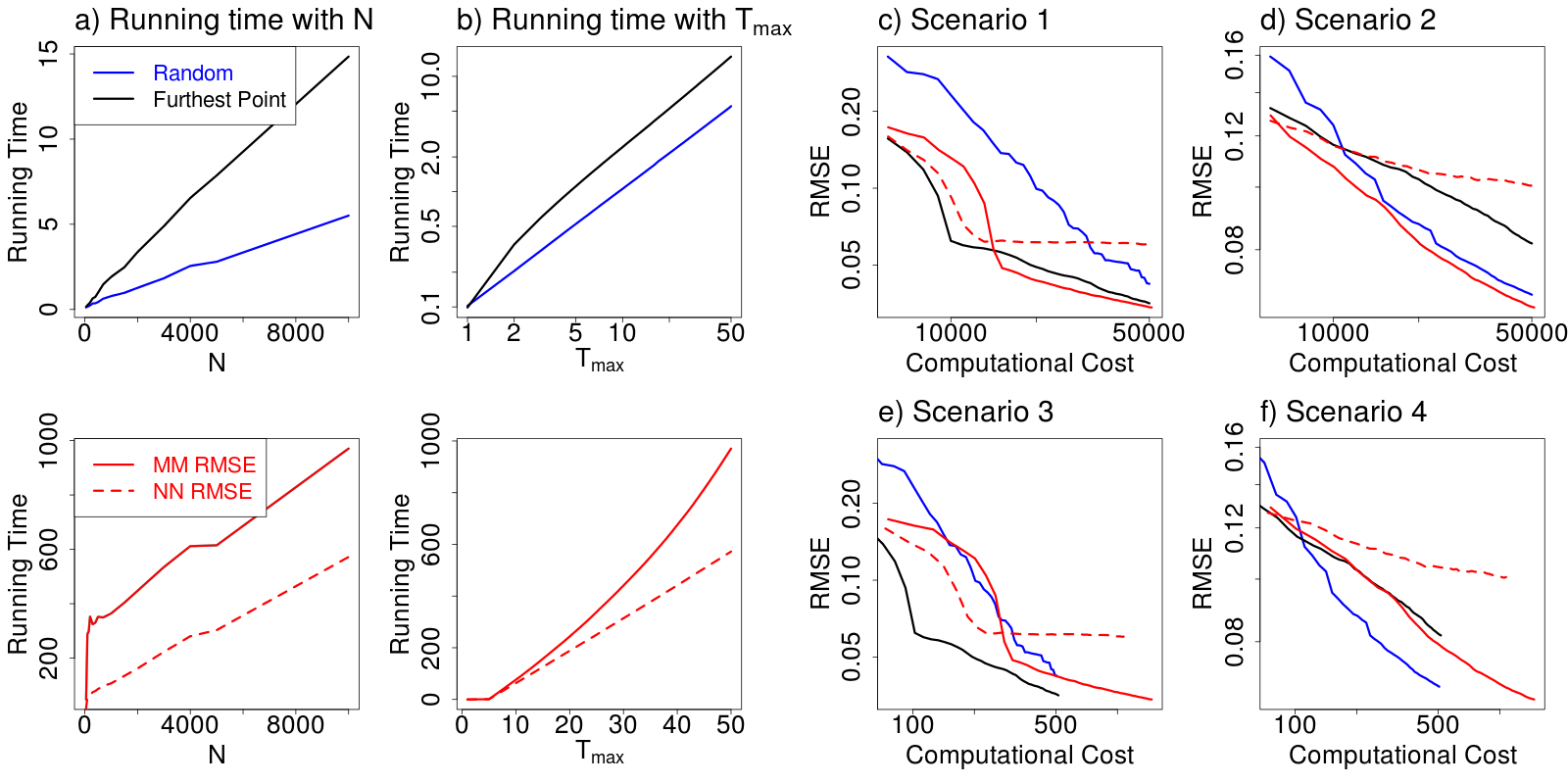}
\caption[] {The scaling and performance considerations of computational cost. a) shows how the running time depends on the matrix size $N$ ($T_{max}=50$) for four configurations of our algorithm. These are the Random Choice Heuristic, the Furthest Distance Heuristic, and the RMSE Loss using either the Mixture Model or Nearest Neighbour emulators.  b) shows the same results for $T_{max}$ (at $N=10000$). Note the different scales between the top and bottom plots. c-f show RMSE as a function of computational time (note the logarithmic x and y axis). This is under different hypothetical scenarios about computational cost and model efficiency (again assuming $N=10000$). Scenarios 1 and 2 assume that $L=0.1$ seconds per matrix element computation, whereas Scenarios 3 and 4 assume that $L=0.001$. Scenarios 1 and 3 use the $\mathrm{RMSE}(t)$ curves from `Dataset~1' (clustered) as shown in Figure~\ref{fig:PredictivePower}, whereas Scenarios 2 and 4 use `Dataset~11' (corrupted). Low RMSE for a given computational cost is desirable; the scenarios have been constructed to illustrate different optimal choices.
\label{fig:scaling} }
\end{centering}
\end{figure}

These arguments are summarised in Table \ref{tab:scaling}, with Figure \ref{fig:scaling}a-b confirming the results. We show the computational cost $C_{choice}$ of using various decision models as a function of either the total number of objects $N$ or the number to be evaluated $T_{max}$. We have constructed the code to make evaluation of $S_{ij}$ approximately negligible so that the scaling of the decision framework dominates. Hence, random decisions take only the amount of time required to define the memory for the data, to copy it, and maintain state (all models share this cost). 

The important criterion is the total compute $C_{choice} + C_{calculation}$, with the cost of evaluating the matrix elements being $C_{calculation} = L T_{max} N$. Hence is it vital to contrast the decision making cost to the calculation cost.  Figure \ref{fig:scaling}c-f show several scenarios of calculation cost whilst varying the problem difficulty. This leads to different optimal choices of decision strategy. Assuming a problem size of $N=10000$, we consider two costs (measured in seconds) per similarity evaluation; $L=0.01$ (Scenarios 1-2) and $L=0.001$ (Scenarios 3-4). The first results in a final cost that is larger the most expensive (RMSE with a mixture model) decision cost by a factor of 5, whereas the second is within the range of computational budgets. By considering the RMSE curves from Figure \ref{fig:PredictivePower}, we can create curves comparing total computational cost to final RMSE, regardless of $T_{max}$. By construction, this leads to a varying optimal decision; when $L$ is large, it pays to use the best model for most values of computational cost. When $L$ is small, it is optimal to either use a cheap heuristic (furthest neighbour) or random samples, depending on their performance in the problem.
 
Figure \ref{fig:scaling}a-b shows that our approach does indeed scale well, with  our implementation in R handling large matrices ($N=10000$ has $100$ million elements) in completely acceptable times. The practical limitation is that we hope to make good decisions, which for large matrices would need heuristics that can be performed in $O(N)$ time or less.  Performing careful calculations on an $O(1)$ fraction of the matrix will not be effective on general similarity matrices; we need a set of heuristics to pull out the most useful objects to consider. Prior information, in the form of covariates or labels, could be very valuable for this. \cite{NIPS2013_4999} use such `side information' in a standard matrix completion framework.

These computational scaling results do not include matrix completion of the final set of objects. This can be performed with any set of chosen objects using any available emulator. The cost of this is $O(N^2 T_{max})$ for nearest neighbour and $O(N^2 T_{max}^2)$ for the mixture model.

Figure \ref{fig:scaling}c-f illustrate several scenarios for which the different computational costs $L$ are accounted for. If $L$ is moderate (Scenarios 1-2, costing $L=0.1$ seconds per matrix element computation) then the RMSE Loss is best. This is universally true for difficult problems whilst the Furthest Distance Criterion is better for small budgets in the clustered dataset, when not much of the matrix is evaluated.  Conversely, if $L$ is very small (Scenarios 3-4 with $L=0.001$ seconds), heuristics are necessary since computing matrix elements is as cheap as emulating them. In Scenario 3 with low $L$ and clustered dataset, the Furthest Distance Heuristic is unbeatable, whereas in Scenario 4 the corrupted dataset means that the Random Choice Heuristic is better for moderate computational budget. 

\section{Genetics example} \label{sec:genetics}

An important application in statistical genetics is the understanding of the relationship between individuals, and the overall structure of that relationship. The process generating genetics data is very well understood - a generative model called the ancestral recombination graph (ARG, \cite{Griffiths1981,Hudson1983}) operates within populations. Between populations, migration occurs at varying rates over time \cite{Lawson2014populations}. Because of this, model-based inference is strongly preferred by this community. However, the true underlying genetic model is prohibitively computationally costly for all but completely unrealistically small and simple samples.  Therefore much effort goes into the development of approximation procedures that capture the key features of the ARG for a manageable computational cost.

We are interested in a population assignment problem \cite{Lawson2012}, for which the likelihood for population assignment can be shown to approximately follow a similarity matrix.  It is therefore of great importance to approximate this matrix.  Fundamentally, the difficulty is that computing the similarity matrix cost $O(N^2L)$ with $L$ taking values up to around 20 million for whole genome resequencing data. The number of individuals sampled $N$ is also growing increasingly large. $N > 1000$ is becoming standard (the first being the 1000 Genome project \cite{10002012integrated}, now with 2500 comparable whole genomes), whilst $N=10000$ lower density samples are in existence (e.g. \cite{ripke2013mega}), and larger datasets are on their way. Computing a row of similarities for this matrix takes days of computation time, and although parallelization is possible, most genetics departments do not have access to the 1000+ core compute farms needed to avoid serial computation. Thus, there is a strong motivation to reduce the computation of these methods, whilst the fundamental computation cannot be changed.  Since $N^2$ and $L$ are of the same order, we will use an $O(NL + N^2)$ implementation of the algorithm.

We apply the method to the well understood Human Genome Diversity Panel dataset formed of $N=938$ individuals from across the globe and $L \approx 2 \times 800,000$ single nucleotide polymorphism (SNP) data points per individual. This is large enough to be difficult to work with but well within the scope of institutional clusters, and was explored in the original \cite{Lawson2012} paper. Our goal here is to recreate the essential features of the matrix at a fraction of the computational cost.  The true rank of this matrix is at least 226 as this is the number of populations identified using full model based inference. This problem contains all of the difficulties of the `hard' simulated data and several more: a) it is not symmetric; b) some individuals are much less similar to any individuals than others; c) it is high rank; d) clusters are of all sizes.

\begin{figure} [ht]
\begin{centering}
\includegraphics[clip,width=1.0\textwidth,angle=0]{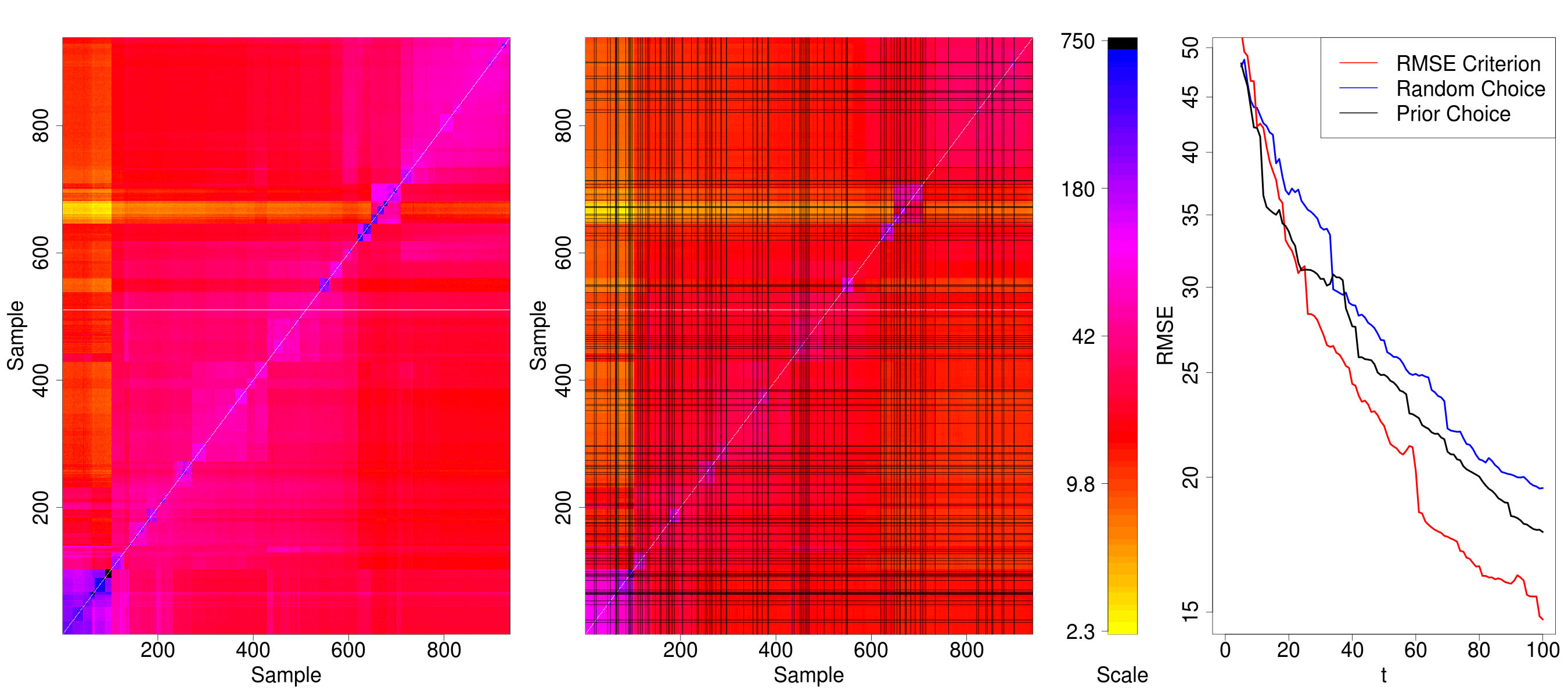}
\caption[] {Genetics example. Left: the raw data ordered as Figure 4A and S14 of \cite{Lawson2012}. Centre: The recovered matrix after 100 iterations, showing the individuals used for the recovery. Right: the root-mean-square-error of the matrix recovery over time, for the RMSE choice criterion, compared to random choice and exploiting the prior information of the 53 population labels by ensuring that each label is sampled in a balanced order. Lower is better.
\label{fig:genetics} }
\end{centering}
\end{figure}

We first converted the genetic similarity to a distance by taking the negative log and adding the new minimum value. We then applied the RMSE Choice criterion with the Mixture Model emulator using cross-validation to estimate $\delta_{i}(t)$. Up to 100 samples were used to recover the matrix.  The performance is shown in Figure \ref{fig:genetics}, compared to both the Random Choice Heuristic and a new `Prior Heuristic'. This exploits prior information about populations using self-declared ethnicities. To do this we sampled one individual from each population, and then a second individual from each population, etc, with both populations and individuals within those populations being defined at random.

\section{Discussion} \label{sec:discussion}

Massive data sets present new challenges for statistical reasoning. The sheer volume of data prevents the routine deployment of even straightforward statistical techniques. Thus a different paradigm is required. Our proposed framework attacks this problem using statistical reasoning to construct a sequential data selection paradigm. At core, this framework acknowledges that it is impossible to complete the statistical evaluation on all data, and instead seeks to find the most useful subset in a sequential manner. This is most suited to problems in which correlation prevents the use of simple sampling, such as the AR model example in Section \ref{sec:simple}, or the central example of similarity matrices. 

The framework combines elements of active learning, modelling, and sequential decision making. At a high-level, the framework features a statistical  emulator (for cheaply estimating unknown data values), a loss, and a choice operator (for selecting among candidates to determine the next computation).  We have provided a number of choices for each, which have different capabilities and computational costs. In the case of similarity matrices, understanding the latent subspace is usually advantageous. However, with massive data this task is difficulty. Hence we have introduced Data Directional Scaling — a method of determining important directions in similarity space itself. This approach allows a rapid characterisation of the main directions without appeal to a costly latent variable model. 

Whilst we explored many advantages of data directional scaling, the key disadvantage is that the dimension of the space grows with the number of samples. It would be fairly simple to extend our decision framework to `discard' directions of the data that are not helpful for prediction. A more sophisticated approach is to perform online clustering to create pseudo-directions via K-means \cite{Hartigan1979} or the more rapid K-medians \cite{jain1988algorithms}. This step brings with it a potential predictive performance drop, since we provide the emulator with less information, and would also need to solve the serious problem of estimating $K$. Further, we cannot compute $\mathcal{S}(i,j)$ for these pseudo-objects although we can still use them in the emulator. Hence we have omitted it from this work, although addressing these concerns is vital to replace the computational scaling factor $T_{max}$ by $K$ in our emulator steps.

We have provided examples in which our approach is to be preferred over matrix completion approaches. This preference is not merely empirical. Matrix completion may still require an intractable computation at large enough data scale, hence precluding its use. Worse, the model assumptions of matrix completion may not match the characteristics of  many similarity matrices, in which cluster structure is often expected. Assessing empirical performance is complicated since we have to handle issues of computational cost, choice function etc. For the type of challenging problem with which we are concerned, where $L$ and $N$ are large, it is important to compare against the benchmark of random selection.  In such cases, empirical results show that the framework is far superior to random selection. 

Further development of this framework will follow in two directions. On the abstract side, we seek some theoretical guarantees on the behaviour of the framework. Moreover, refinements of the details of the framework, emulator and choice operator, should lead to performance enhancements. On the computational side, extension of the approach to parallel and cloud systems is of great interest. The big data we are interested in will usually reside on a distributed storing system, such as HADOOP.  

\bibliographystyle{mychicago}

\end{document}